\documentclass[10pt]{wlscirep}
\usepackage[utf8]{inputenc}
\usepackage[T1]{fontenc}
\title{A simple iterative map forecast of the COVID-19 pandemic}
\author[1,*]{Andr\'{e} E. Botha}
\author[2]{Wynand Dednam}
\affil[1]{Department of Physics, University of South Africa, Private Bag X6, Florida 1710, South Africa}
\affil[2]{Departamento de Fisica Aplicada, Universidad de Alicante, San Vicente del Raspeig, 03690 Alicante, Spain}
\affil[*]{bothaae@unisa.ac.za}

\begin{abstract}
We develop a simple 3-dimensional iterative map model to forecast the global spread of the coronavirus disease. Our model contains at most two fitting parameters, which we determine from the data supplied by the world health organisation for the total number of cases and new cases each day. We find that our model provides a surprisingly good fit to the currently-available data, which exhibits a cross-over from exponential to power-law growth, as lock-down measures begin to take effect. Before these measures, our model predicts exponential growth from day 30 to 69, starting from the date on which the world health organisation provided the first `Situation report' (21 January 2020 -- day 1). Based on this initial data the disease may be expected to infect approximately 23\% of the global population, i.e. about 1.76 billion people, taking approximately 83 million lives. Under this scenario, the global number of new cases is predicted to peak on day 133 (about the middle of May 2020), with an estimated 60 million new cases per day. If current lock-down measures can be maintained, our model predicts power law growth from day 69 onward. Such growth is comparatively slow and would have to continue for several decades before a sufficient number of people (at least 23\% of the global population) have developed immunity to the disease through being infected. Lock-down measures appear to be very effective in postponing the unimaginably large peak in the daily number of new cases that would occur in the absence of any interventions. However, should these measure be relaxed, the spread of the disease will most likely revert back to its original exponential growth pattern. As such, the duration and severity of the lock-down measures should be carefully timed against their potentially devastating impact on the world economy. 
\end{abstract}
\begin{document}

\flushbottom
\maketitle

\thispagestyle{empty}

\section*{Introduction}
On 11 March 2020, the world health organisation (WHO) characterised the 2019 outbreak of coronavirus disease (COVID-19) as a pandemic, referring to its prevalence throughout the whole world\cite{whostate}. The outbreak started as a pneumonia of an unknown cause, which was first detected in the city of Wuhan, China. It was reported as such to the WHO on the 31st December 2019, and has since reached epidemic proportions within China, where it has infected more than 80 000 citizens, to date. During the first six weeks of 2020 the disease spread to more than 140 other countries, creating wide-spread political and economic turmoil, due to unprecedented levels of spread and severity.  

The rapid spread of COVID-19 is fuelled by the fact that the majority of infected people do not experience severe symptoms, thus making it more likely for them to remain mobile, and hence to infect others\cite{qiu20}. At the same time the disease can be lethal to some members of the population, having a globally averaged fatality ratio of about $6\%$, at present (16 April 2020). Furthermore the new virus seems capable of surviving for unusually long periods on plastic and metal surfaces, which are both, frequently encountered in everyday life. Detailed analysis of the virus also revealed that its outer surface consists of club-like “spikes” that are about four times more effective at establishing the infection than in the closely related coronavirus that causes severe acute respiratory syndrome (SARS) in 2002-2003\cite{sha20}.  It is most likely this particular combination of traits that has made the COVID-19 outbreak one of the largest in recorded history. 

While there are a number of models available for the global spread of infectious diseases\cite{rab20}, some even containing very sophisticated traffic layers\cite{gleamviz}, relatively few researchers are making use of simpler models that can provide the big picture without being difficult to interpret unambiguously. In the latter category of relatively simple models we could find only a discrete epidemic model for SARS~\cite{zho09}, and more recently, a comparison of the logistic growth and susceptible-infected-recovered (SIR) models for COVID-19\cite{bat20}.

In our present work we develop a simple discrete 3-dimensional iterative map model, which shares some similarities with the classic SIR model. We show that our model can fit both the initial exponential growth in the number of cases, before lock-down measures began to take effect (about day 69), as well as subsequent power-law growth. In our view the current interventions are necessary to prevent the unimaginably fast (exponential) spread of the disease until some solution (perhaps in the form of a vaccine) can be developed. 

\section*{Results}
\subsection*{Simple model}
As a simple exponential growth model for the global data we initially develop a 3-dimensional iterative map model given by
\begin{eqnarray}
x_{i+1}&=& x_i+\alpha y_i\left(z_{0} - x_i\right)/z_{0}  \nonumber  \\
y_{i+1} &=&  \alpha y_i\left(z_{0} - x_i\right)/z_{0} \label{eq1} \\
z_{i+1} & = &  z_i - cy_i \nonumber ,
\end{eqnarray}
where $x_i$ is the total number of confirmed cases, $y_i$ is the number of new cases and $z_i$ is the global population, on any given day $i$. We denote the only fitting parameter by $\alpha$, while $c$ is a fixed parameter equal to the fraction of people who have died from the disease. According to the latest available global data from the WHO (see the last row of  Table~\ref{tab1} in {\bf Methods} ), $c = 0.06343$. 

We briefly describe the physical content of Eqs.~(\ref{eq1}). The first equation simply updates the total number of cases by setting it equal to the previous total number of cases, plus the number of new cases. Here the factor of $1/z_{0}$ has been introduced for convenience, to ensure that the proportionality constant $\alpha$ remains close to unity. In the second equation see that the number of new cases is assumed to be proportional to the previous number of new cases multiplied by the previous number of susceptible people. For simplicity we assume here the the number of susceptible people are all those who have not yet had the disease. The third equation is merely to keep track of the global population, by subtracting the estimated number of people who have died each day, based on the fraction $c$.

We have found that our simple model provides a very good fit to the global data between the 29th and 70th days of the pandemic. By using Levenberg--Marquardt (least squares) optimisation\cite{mar63} (see {\bf Methods}), we find $\alpha = 1.1393$, for the initial condition $x_{30} = 75194.3 $ $y_{30} = 349.477$ and $z_{30} = z_{0} = 7.7000\times 10^{9}$. 

\begin{figure}[ht!]
\centering
\includegraphics[width=0.49\linewidth]{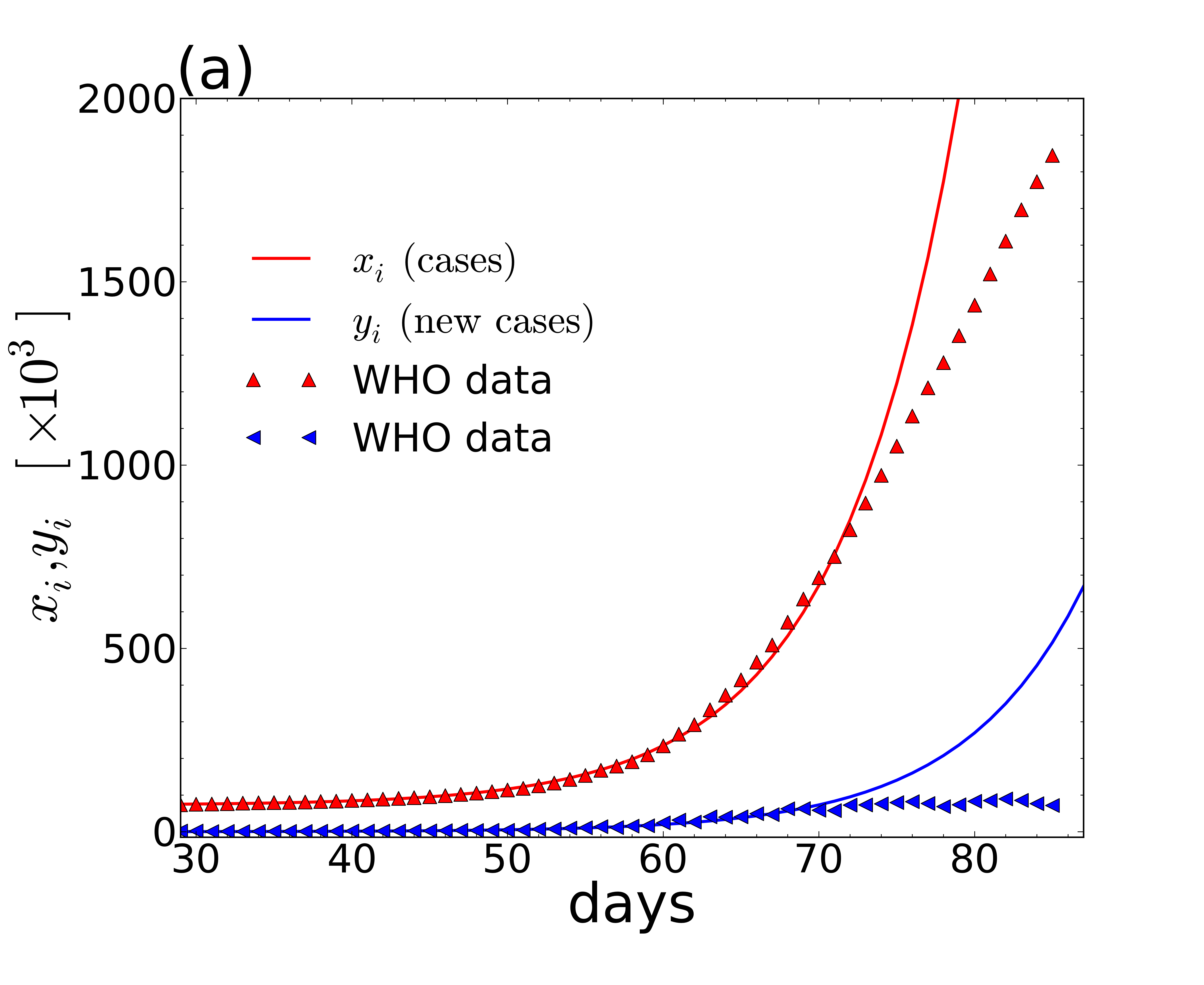}
\includegraphics[width=0.49\linewidth]{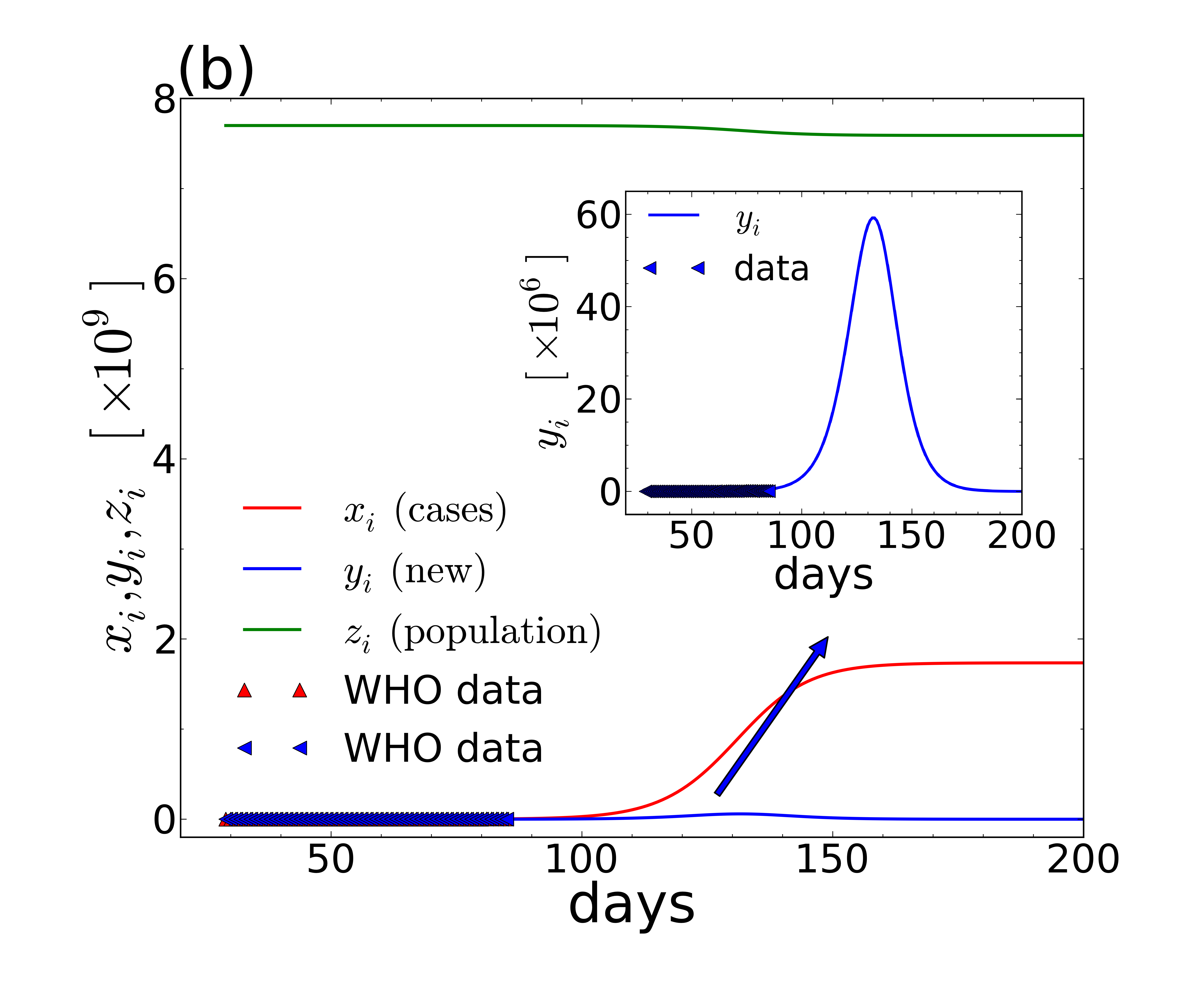}
\caption{A comparison of the data with Eqs.~(\ref{eq1}). (a) Close up view of the extent to which the model fits the global data for the total number of confirmed cases $x_i$ and new cases $y_i$, so far. (b) A forecast of how the disease is likely to spread globally, up to the 200th day, i.e. 7 August 2020. As the inset shows, the peak number of new infections is predicted to occur on day 133 (19 May 2020), when about 60 million new cases may be expected.}
\label{fig1}
\end{figure}

Figure~\ref{fig1} shows a comparison of the data with the model, as well as a forecast made up to the 200th day.
As we see in Figure~\ref{fig1}(a), the model provides a good fit to the data between the 29th and 70th days, having a correlation coefficient (R-squared value) of 0.99432 or higher (see Table~\ref{tab1}). The forecast made in Figure~\ref{fig1}(b) (corresponding to the 69th 'Last day' in Table~\ref{tab1}), predicts that approximately a quarter of the worlds population, i.e. $\approx 1.76/7.7=0.23$, would have had COVID-19 by the 200th day. The peak of the pandemic is expected to occur on day 133, when about 60 million daily new cases can be expected. We also predict that by the beginning of August 2020, hardly any new cases should occur; however, the total number of lives lost by then could be as high as 83 million, i.e. a number which is almost the same as the global population increase for the year 2019\cite{worldometer}.

\begin{figure}[ht!]
\centering
\includegraphics[width=0.49\linewidth]{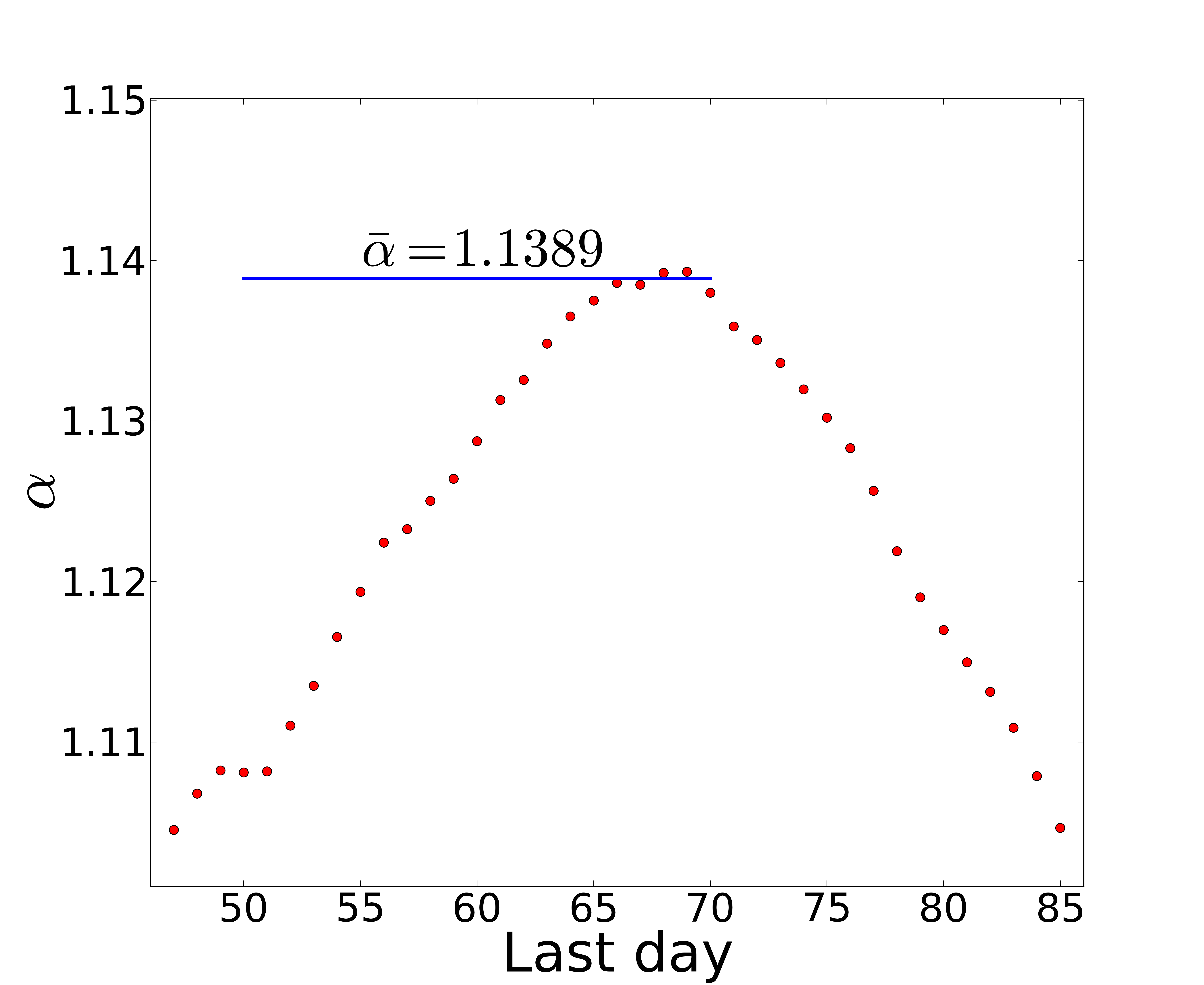}
\caption{Variation of the fitting parameter $\alpha$ as more and more of the available data is used in the fitting procedure. We see that the value of $\alpha$ reaches its maximum on 'Last day' 69. The average value of $\alpha$ over 'Last day' 66-69 is given by $\bar{\alpha} = 1.1389$.}
\label{fig2}
\end{figure}

\begin{table}[htb!]
\centering
\begin{tabular}{|c|c|c|c|c|c|c|c|}
\hline
Last day & $\alpha$  & $x_{200}\times10^9$ & $\max \{y\}\times10^6$& Day of max$\{y\}$ & Deaths$\times10^6$ & $c$ &  R-squared \\
\hline \hline
59 & 1.12640 & 1.628 & 50.487 &  140 & 68.087 & 0.04183 & 0.99880 \\
60 & 1.12874 & 1.653 & 52.218 &  138 & 69.506 & 0.04204 & 0.99833 \\
61 & 1.13131 & 1.682 & 54.179 &  137 & 70.702 & 0.04203 & 0.99684 \\
62 & 1.13256 & 1.693 & 55.054 &  136 & 74.092 & 0.04376 & 0.99615 \\
63 & 1.13483 & 1.719 & 56.817 &  135 & 74.895 & 0.04358 & 0.99478 \\
64 & 1.13652 & 1.737 & 58.143 &  134 & 75.637 & 0.04354 & 0.99396 \\
65 & 1.13751 & 1.746 & 58.804 &  134 & 77.747 & 0.04452 & 0.99347 \\
66 & 1.13861 & 1.758 & 59.711 &  133 & 79.136 & 0.04503 & 0.99361 \\
67 & 1.13849 & 1.755 & 59.571 &  133 & 80.423 & 0.04583 & 0.99318 \\
68 & 1.13923 & 1.762 & 60.100 &  133 & 81.663 & 0.04634 & 0.99402 \\
69 & 1.13930 & 1.761 & 60.097 &  133 & 83.118 & 0.04719 & 0.99432 \\
70 & 1.13799 & 1.746 & 59.065 &  133 & 83.399 & 0.04776 & 0.99280 \\
71 & 1.13589 & 1.722 & 57.375 &  134 & 83.498 & 0.04848 & 0.98965 \\
72 & 1.13505 & 1.712 & 56.673 &  135 & 84.377 & 0.04929 & 0.98965 \\
73 & 1.13362 & 1.694 & 55.490 &  135 & 86.020 & 0.05078 & 0.98828 \\
74 & 1.13197 & 1.674 & 54.213 &  136 & 86.655 & 0.05175 & 0.98633 \\
75 & 1.13021 & 1.651 & 52.778 &  137 & 89.481 & 0.05419 & 0.98396 \\
76 & 1.12831 & 1.629 & 51.307 &  138 & 90.189 & 0.05538 & 0.98096 \\
77 & 1.12565 & 1.599 & 49.314 &  140 & 89.228 & 0.05582 & 0.97413 \\
78 & 1.12189 & 1.556 & 46.592 &  142 & 88.258 & 0.05674 & 0.95830 \\
79 & 1.11901 & 1.521 & 44.480 &  144 & 89.030 & 0.05855 & 0.94451 \\
\hline
\end{tabular}
\caption{\label{tab1} Variation in the predictions made by the model using only data up to and including the 'Last day', as indicated in the first column. The predicted total number of 'Deaths' indicated in column 6 is equal to $z_{30} - z_{200}$, i.e. the difference between the initial population and the remaining population on day 200. }
\end{table}
In Table~\ref{tab1} we see that the fitting parameter $\alpha$, and hence the predictions made by the model, do change somewhat as more of the available data is used in the fitting procedure. To see the variation in $\alpha$ more clearly we have plotted the first and second columns of Table~\ref{tab1} in Figure~\ref{fig2}. It shows that, as more data is used, there seems to be a general upward trend in $\alpha$, until the 69th 'Last day'. Beyond this our 'Simple model' no longer provides a good fit to the data, as is evident from the rapidly declining R-squared values that are given in the far right column of the Table.  

In Figure~\ref{fig2} we also plot (blue solid line) the mean value $\bar{\alpha} = 1.1389$ over 'Last day' 66-69. The value of $\alpha$ is essentially constant over these four days. 

\subsection*{Transition from exponential to power-law growth}
China is currently the only country in which the rapid further spread of the virus appears to be under control. An investigation of the data for China observed that the total number of infections as well as the number of recoveries and the number of deaths followed a power law growth, rather than growing exponentially\cite{Ziff2020,li2020scaling}. A wider study for 25 different countries found that other countries also showed scale free power law growth behaviour with different country-specific scaling exponents\cite{singer2020shortterm}. Subsequent in-depth analysis of more available data of country-specific growth behaviours revealed, however, that different countries exhibit generally different growth behaviours\cite{singer2020Covid19pandemic}. It was found that while indeed some countries such as for example the Netherlands or Norway exhibit a power law behaviour over nearly the whole range of the available data, other countries such as for example the US or France display long periods of very few infections followed by a massive exponential surge but subsequently quickly reduce the growth rate again: a behaviour that is clearly not describable by a power law. In general, however, the spread of the disease within individual countries initially appears to be exponential, followed by power law growth once national lock-down measures take effect. We may therefore expect a similar trend in the global data, which is more or less what we observe from day 70 onward. Note that day 70 corresponds to the 30th of March 2020, which is two to three weeks after many of the hardest hit European countries implemented their national lock-downs: Italy (10 March), Spain (14 March), France (17 March), etc.

One can rationalise the transition from exponential to power law growth as follows. Any person living his/her life in the city has on any given day a large number of contacts with people and infectious objects. Of these interactions, relatively few can be classified as relations that are of particular meaning to that individual, such as contact with a spouse, children, friends, family relations, coworkers, etc. Thus most contacts are completely random interactions, such as through public transport, contacts with clients, public parks, shops, entertainment facilities, etc. For the spread of the virus it is irrelevant whether the susceptible person's interaction with another individual is meaningful or random, as long as the virus can be transmitted.

Since random interactions are generally never with the same people or infectious objects, they generally give rise to an exponential spread of the disease. The introduction of governmental measures, such as a 'lock-down', severely reduce random interactions whilst leaving more open the possibility of relationship interactions. Human societies (friendships, relations, collaborations, social networks etc.) have been found to be organised in small world or scale-free networks\cite{Watts1998,Watts2004} exhibiting power law behaviours. Given that the further spreading of the epidemic is now effectively being restricted, on a global scale, to only those relation networks, it follows that the global spread pattern should also change from exponential to a power law growth, as has been shown in countries such as China, the Netherlands, Norway and others.

In light of this expectation we have expanded our original simple model to reflect the change of the growth behaviour after world wide governmentally instituted lock-down measures have started to take effect, i.e.on about day 69, according to the global data.

\subsection*{Improved model}
As described in {\bf Methods} we have derived an improved model which can take into account both exponential and power law growth. This model is given by   
\begin{eqnarray}
x^{\prime}_{i+1}&=& x_{i}^{\prime} + y_i^{\prime} \exp \left( b \left[ \frac{a}{y_{i}^{\prime}+a} \right]^{\frac{1}{b}} \right) \left(z_{i}^{\prime} - x_{i}^{\prime}\right)/z_0  
\nonumber  \\
y^{\prime}_{i+1} &=&  y^{\prime}_i \exp \left( b \left[ \frac{a}{y^{\prime}_i+a} \right]^{\frac{1}{b}} \right) \left(z^{\prime}_i - x^{\prime}_i\right)/z_0 \label{eq2} \\
z^{\prime}_{i+1} & = &  z^{\prime}_i - cy^{\prime}_i \nonumber ,
\end{eqnarray}
where there are now two fitting parameters, $a$ and $b$.

Equation~(\ref{eq2}) is more general than (\ref{eq1}). In the limit, $a >> \max\{y^{\prime}_i \}$, it reduces to the same form as equation~(\ref{eq1}), since the exponential factor essentially becomes independent of $y^{\prime}_i$, i.e. $a/(y^{\prime}_i+a) \xrightarrow[\infty]{a} 1$ and, by comparison to equation~(\ref{eq1}), $\alpha = \exp \left( b \right)$.

The main advantage of equation~(\ref{eq2}) is that it can take into account the observed cross-over from exponential growth during the initial phase of the pandemic, to a power law growth, now that measures to curb the spread are starting to take effect on a global scale. 

\begin{figure}[ht!]
\centering
\includegraphics[width=0.49\linewidth]{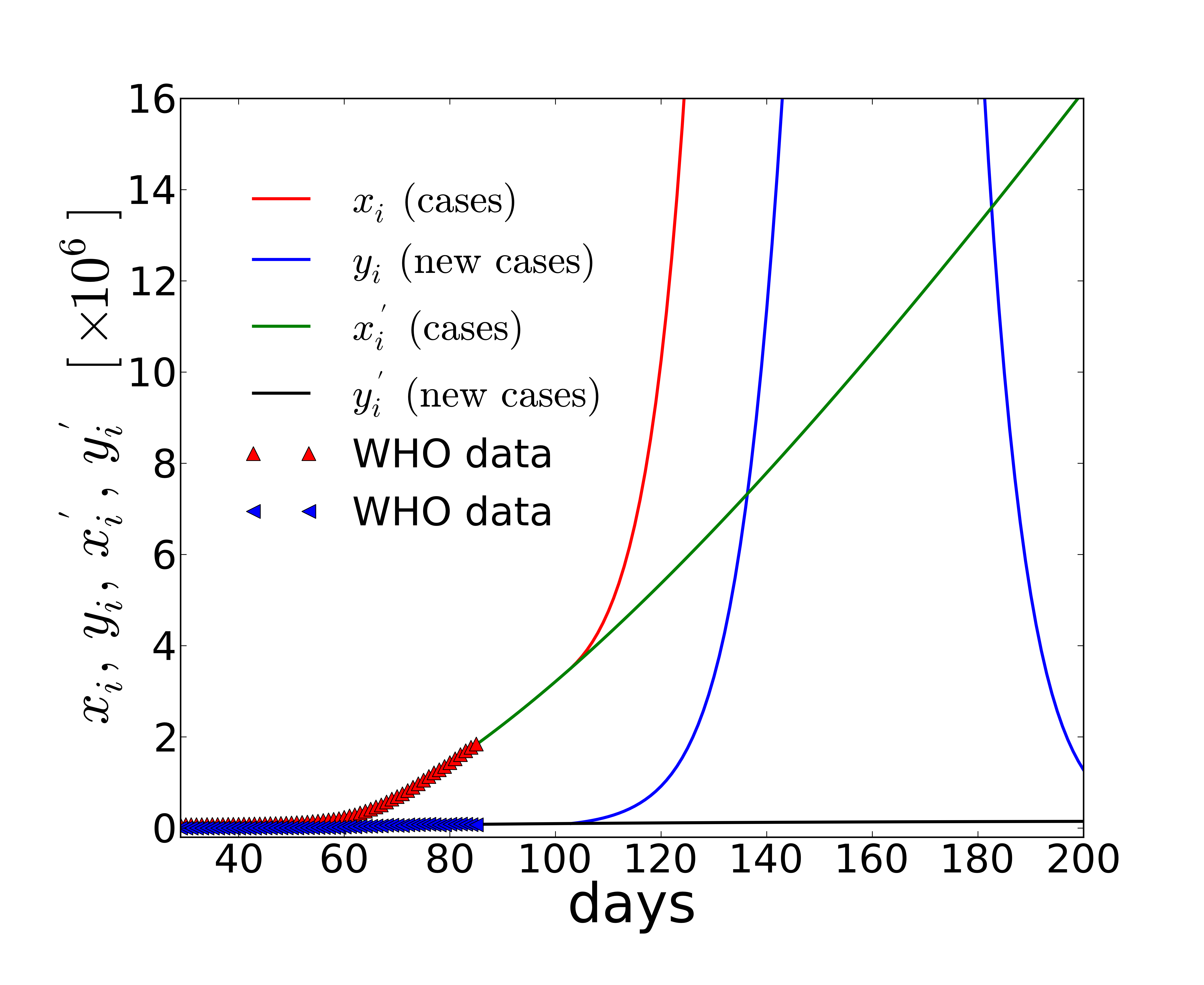}
\caption{A comparison of the forecasts given by the 'Simple' (red and blue curves) and 'Improved' (green and black curves) models. With the 'Improved' model (Eqn. (\ref{eq2}) with the fitting parameters $a = 85710$ and $b = 0.23756$) we find that the growth in the number of cases increases according to a power law with exponent less than 1. However, should all lock-down measures be lifted on day 100 (i.e. the by the end of April 2020), one can expect the growth to revert back to its original exponential behaviour (Eqn. (\ref{eq1}) with $\alpha = 1.1393$). }
\label{fig3}
\end{figure}

Since our 'Improved model' can take into account the effects of the interventions put in place to curb the global spread of the virus, we were expecting it to provide a more optimistic forecast than that of our 'Simple model'. However, although our 'Improved model' does predict a vast reduction in the daily number of new cases, it also predicts a prolonged spread of the disease, as long as the interventions are strictly enforced. This is shown by the green curve in Figure~\ref{fig3}.  
The problem however is that this type of power-law growth is in fact so gradual, that it would have to be maintained for several decades before a significant number of the population become immune to the disease, i.e. as a result of having contracted the virus. At the same time, should the stringency of the interventions be relaxed at any time, the spread will revert to its original exponential growth behaviour, implying that the spread of the disease could quickly get out of hand, again and again. In Figure~\ref{fig3} we assume, for example, that all restrictions are lifted at the end of April (day 100), giving rise to the exponential increase shown by the red curve. The blue curve shows the daily number of new cases undergoes hardly any reduction in its maximum peak height, though its peak is shifted forward by about a month, in comparison to the forecast given in Figure~\ref{fig1}.     

\section*{Discussion}
One can of course try to answer more specific questions with more sophisticated models, like the discrete model we mentioned for SARS~\cite{zho09}; however, here we have been more interested in developing a very simple model that brushes over the details and only captures the essential, large scale behaviour. 

As we have seen, there are essentially two possible behaviours: either the virus can proceed naturally, leading to exponential growth in the number of cases; or else its progression can be slowed down to a more manageable power-law growth, i.e. for as long as the present, stringent, lock-down measures can be imposed. But for how long will it be necessary (or possible) to impose the present measures? On the near horizon there does not seem to be any easy answer to this question, nor is there any clear solution to the pandemic itself. A vaccine is unlikely to be developed for several months, and there seems to be only a remote possibility that the virus may mutate naturally into something less lethal to humans.

The direct human cost of an unchecked (exponential) spread of the virus could be truly devastating\cite{nat20}. On the one hand, it could result in a catastrophic loss of tens of millions of lives, as our model predicts, but on the other hand it will end the pandemic once and for all. The current measures, leading to more manageable power-law growth, come at a very high economic cost and are therefore not sustainable for much longer. So far these measures have included enforced quarantine, which has led to a severe slowdown in economic activity and manufacturing production, principally due to declining consumption and disrupted global supply chains\cite{supplychain}. (As an example of the severity of the slowdown in production, several major car manufacturers are gradually halting production in major manufacturing hubs throughout the developed world \cite{carhalt}.) This decline, coupled with the associated economic uncertainty, has had knock on effects in the form of historically unprecedented stock market falls\cite{stockcrash}. Although the stock market is more of an indicator of the future value of the profits of listed corporations, their collapsed share prices could trigger severe financial crises because of a spike in bankruptcies. (The debt of US corporations is the highest it has ever been\cite{modernjubilee}.) The inevitable loss of jobs will also lead to an inability to pay bills and mortgages, increased levels of crime, etc. In principle, such a major decline in economic conditions could also result in an equally large-scale loss of life/livelihoods, albeit over a more prolonged period of time.

\section*{Methods}
\subsection*{The global WHO data}
We have fitted our models to the global data extracted from the daily `Situation reports' made available through the world health organisation's webpage~\cite{whodata}. For convenience we reproduce this data in Table~\ref{tab2}.
\begin{table}[htb!]
\centering
\begin{tabular}{|l|l|l|l|}
\hline
Day & Cases & New & Deaths \\
\hline
30 & 75204  & 1872  & 2009 \\
31 & 75748  & 548   & 2129 \\
32 & 76769  & 1021  & 2247 \\
33 & 77794  & 599   & 2359 \\
34 & 78811  & 1017  & 2462 \\
35 & 79331  & 715   & 2618 \\
36 & 80239  & 908   & 2700 \\
37 & 81109  & 871   & 2762 \\
38 & 82294  & 1185  & 2804 \\
39 & 83652  & 1358  & 2858 \\
40 & 85403  & 1753  & 2924 \\
41 & 87137  & 1739  & 2977 \\
42 & 88948  & 1806  & 3043 \\
43 & 90869  & 1922  & 3112 \\
44 & 93091  & 2223  & 3198 \\
45 & 95324  & 2232   & 3281 \\
46 & 98192  & 2873   & 3380 \\
47 & 101927 & 3735   & 3486 \\
48 & 105586 & 3656   & 3584 \\
49 & 109577 & 3993   & 3809 \\
\hline
\end{tabular}
\hspace*{0.5cm}
\begin{tabular}{|l|l|l|l|}
\hline
Day & Cases & New  & Deaths\\
\hline
50 & 113702 & 4125   &  4012 \\
51 & 118319 & 4620   &  4292 \\
52 & 125260 & 6741   &  4613 \\
53 & 132758 & 7499   &  4955 \\
54 & 142539 & 9764   &  5392 \\
55 & 153517 & 10982  &  5735 \\
56 & 167511 & 13903  &  6606 \\
57 & 179112 & 11526  &  7426 \\
58 & 191127 & 15123  &  7807 \\
59 & 209839 & 16556  &  8778 \\
60 &  234073 & 24247  &  9840 \\
61 &  266073 & 32000  & 11184 \\
62 &  292142 & 26069  & 12784 \\
63 &  332930 & 40788  & 14510 \\
64 &  372757 & 39827  & 16231 \\
65 &  414179 & 40712  & 18440 \\
66 &  462684 & 49219  & 20834 \\
67 &  509164 & 46484  & 23335 \\
68 &  571678 & 62514  & 26494 \\
69 &  634835 & 63159  & 29957 \\
\hline
\end{tabular}
\hspace*{0.5cm}
\begin{tabular}{|l|l|l|l|}
\hline
Day & Cases & New  & Deaths\\
\hline
70 &  693224 & 58411  & 33106 \\
71 &  750890 & 57610  &  36405 \\
72 &  823626 & 72736  &  40598 \\
73 &  896450 & 72839  &  45525 \\
74 &  972303 & 75853  &  50321 \\
75 & 1051635 & 79332  &  56985 \\
76 & 1133758 & 82061  &  62784 \\
77 & 1133758 & 82061  &  62784 \\
78 & 1279722 & 68766  &  72614  \\
79 & 1353361 & 73639  &  79235  \\
80 & 1436198 & 82837  &  85522  \\
81 & 1521252 & 85054  &  92798  \\
82 & 1610909 & 89657  &  99690 \\
83 & 1696588 & 85679  & 105952 \\
84 & 1773084 & 76498  & 111652 \\
85 & 1844863 & 71779  & 117021 \\
\hline
\end{tabular}
\caption{\label{tab2} Data used for the Levenberg--Marquardt (least squares) optimisation of the parameter $\alpha$ in Eqs.~\ref{eq1}. The columns contain the day since the first situation report (21 January 2020), the total number of cases (column 2), the number of new cases (column 3), and the total number of deaths (column 4).}
\end{table}

\subsection*{Python script for fitting the parameters}
For the reader's convenience, the complete python script for the optimisation of the single parameter $\alpha$ of the simple model is provided on the following page. In this script, the function \verb|leastsq()|, imported from the module
 \verb|scipy.optimize|\cite{lan08}, uses Levenberg--Marquardt optimization to minimize the residual vector returned by the function \verb|ef()|. The function \verb|leastsq()| is called from within \verb|main()|, which reads in the data and sets up the initial parameter and the other two quantities (the initial values \verb|x[0]| and \verb|y[0]|) for optimisation. These three quantities are then passed to \verb|leastsq()|, via the vector \verb|v0|.

For the data in Table~\ref{tab2}, the output from the script should be:
\begin{verbatim}
R-squared = 0.99432
alpha = 1.1393
x[29] = 75194.3
y[29] = 349.477
z[29] = 7700000000.0
\end{verbatim}
The python scripts that were used to produce the Figures and Table 1 in this paper are available from the corresponding author upon request.

\newpage
\begin{verbatim}
from scipy import linspace,sqrt,dot,concatenate,zeros,loadtxt
from scipy.optimize import leastsq
#
def ef(v,x,y,z,c,xdata,ydata):
    '''
    # Residual (error function)
    '''
    x[0] = v[1]; y[0] = v[2]
    for i in range(len(xdata)):
        y[i+1] = v[0]*y[i]*(z[i]-x[i])/z[0]
        x[i+1] = x[i] + y[i+1]
        z[i+1] = z[i] - c*y[i]
    er = concatenate(((x[0:-1]-xdata)/xdata,(y[0:-1]-ydata)/ydata))
    return er
#
def main():
    f = loadtxt('data.dat')     # (WHO data as in Table 1)
    xdata = f[0:41,1]
    ydata = f[0:41,2]
    c = f[-1,3]/f[-1,1]         # fraction of deaths occuring
    n = len(xdata)+1            # number of data points
    m = int(f[0,0])             # day of first data point
    #
    x = zeros(n,'d')            # for storing total cases
    y = zeros(n,'d')            # for storing new cases
    z = zeros(n,'d')            # for storing total population
    #  
    x[0] = 76000                # initial guess -- to be optimized
    y[0] = 300                  # initial guess -- to be optimized
    z[0] = 7.7e9                # estimated total global population
    #
    v0 = zeros(3,'d')           # quantities to be be optimised    
    v0[0] = 1.14                # initial guess for alpha 
    v0[1] = x[0]
    v0[2] = y[0]
    #
    v,cov,infodict,mesg,ier = leastsq(ef,v0,args=(x[0:n],y[0:n],z[0:n],c,xdata
                               ,ydata),full_output=True,ftol=1e-13,maxfev=500)
    #
    ys = concatenate((x[0:n-1], y[0:n-1]))
    er = concatenate(((x[0:-1]-xdata),(y[0:-1]-ydata)))
    ssErr = dot(er,er)
    ssTot = ((ys-ys.mean())**2).sum()
    rsquared = 1.0-(ssErr/ssTot )
    x[m] = v[1]; y[m] = v[2]; z[m] = 7.7e9
    print('R-squared = ' + str(round(rsquared,5)))
    print('alpha = ' + str(round(v[0],5)))
    print('x['+str(m)+'] = ' + str(round(x[m],1)))
    print('y['+str(m)+'] = ' + str(round(y[m],3)))
    print('z['+str(m)+'] = ' + str(round(z[m],1)))

if __name__ == "__main__":
   main()

\end{verbatim}

\subsection*{Improved model equations}
The form of the equations in our improved model was motivated by a Taylor series expansion of the power law 
\begin{equation}
y\left( t\right) = at^{b}\text{, } \label{eq3}
\end{equation} 
where $a$ and $b$ are real positive constants. For convenience we first take the natural logarithm on both sides of (\ref{eq3}), 
\begin{equation}
\ln y\left( t\right) =\ln a+b\ln \left( t\right) \text{.}
\label{eq4}
\end{equation} 
For $\Delta t\ll t$, we then expand about $t$, to get  
\begin{eqnarray}
\ln y\left( t+\Delta t\right)  &=&\ln a+b\ln \left( t+\Delta t\right)  
\nonumber \\
&\approx &\ln a+b\left( \ln t+\frac{\Delta t}{t}-\frac{1}{2}\left( \frac{ 
\Delta t}{t}\right) ^{2}+\frac{1}{3}\left( \frac{\Delta t}{t}\right)
^{3}-\ldots \right) \text{.} \label{eq5}
\end{eqnarray} 
Subtracting (\ref{eq4}) from (\ref{eq5}) produces, to leading order in $\Delta t/t$, 
\begin{equation}
\ln \left( \frac{y\left( t+\Delta t\right) }{y\left( t\right) }\right)
\approx b\frac{\Delta t}{t}\text{.} \label{eq6}
\end{equation} 
From (\ref{eq4}) we also see that 
\begin{equation}
\ln \left( t\right) =\frac{\ln y\left( t\right) -\ln a}{b}=\ln \left( \frac{ 
y\left( t\right) }{a}\right) ^{\frac{1}{b}}\text{,}
\label{eq7}
\end{equation} 
or 
\begin{equation}
\frac{1}{t}=\left( \frac{a}{y\left( t\right) }\right) ^{\frac{1}{b}}\text{.}
\label{eq8}
\end{equation} 
After substituting (\ref{eq8}) into (\ref{eq6}), and taking the exponential on both sides of the resulting equation, we arrive at
\begin{equation}
y\left( t+\Delta t\right) \approx y\left( t\right) \exp \left( b\Delta
t\left( \frac{a}{y\left( t\right) }\right) ^{\frac{1}{b}}\right) \label{eq9} 
\end{equation} 
In terms of an iterative map the power law growth given by (\ref{eq9}) suggests the form 
\begin{equation}
y_{i+1}=y_{i}\exp \left( b\left[ \frac{a}{y_{i}+a}\right] ^{\frac{1}{b} 
}\right) \label{eq10}
\end{equation}
where we have put $\Delta t=1$ and added a shift in the denominator ($ y_{i}\rightarrow y_{i}+a$) for convenience. This shift allows the iterative map model to change smoothly from pure exponential growth ($ y_i << a $), to something that approximates power law growth for values of $y_i$ that become comparable to $a$. Note that, because of the latter approximation, the (now) fitting parameter $b$ is no longer equal to the power indicated by $b$ in Eqn.~(\ref{eq3}).

\bibliography{covid19}

\section*{Acknowledgements}
A. E. B. would like to acknowledge M. R. Kolahchi, H. M. Singer and V. Hajnov\'{a} for helpful discussions about this work. Both authors wish to thank A. Thomas for uncovering some of the related literature.

\section*{Author contributions statement}
A. E. B devised the research project, developed the models, performed the numerical simulations and produced all the figures. Both authors analysed the results and wrote the paper.

\section*{Ethics declarations}
The authors declare no competing interests.

\end{document}